\documentclass[aip, reprint, nofootinbib]{revtex4-1}
\usepackage{graphicx}

\usepackage[toc, section=section, nogroupskip, acronym]{glossaries}

\usepackage{amsmath}
\usepackage{amssymb,amsmath}
\usepackage[usenames,dvipsnames]{color}
\usepackage{cancel}
\usepackage[margin=.7in]{geometry}
\usepackage{braket}
\usepackage{enumitem}
\usepackage{mathtools}
\usepackage{setspace}
\usepackage[english]{babel} 
\usepackage[titletoc]{appendix}
\usepackage{titlesec}
\usepackage{hyperref}
\usepackage{ntheorem}
\theoremseparator{:}
\usepackage{ulem}

\usepackage{acro}
\usepackage[nottoc]{tocbibind}

\pdfoutput=1

\begin{document}

\title{Importance of the `Higgs' Amplitude Mode in Understanding the ``Ideal Glass Transition'' and the Kauzmann Entropy Paradox at the Four-Dimensional Crystal/Glass Quantum Critical Point}

\author{Caroline S. Gorham}
\email{caroling@cmu.edu}
\affiliation{Department of Materials Science and Engineering, Carnegie Mellon University, Pittsburgh, PA 15213, USA}

\author{David E. Laughlin}
\email{laughlin@cmu.edu}
\affiliation{Department of Materials Science and Engineering, Carnegie Mellon University, Pittsburgh, PA 15213, USA}

\begin{abstract}
In this article, a theoretical description of the ``ideal glass transition'' is approached upon the adoption of a quaternion orientational order parameter. Unlike first-order phase transitions of liquids into crystalline solid states, glass transitions are entirely different phenomena that are non-equilibrium and that are highly dependent on the applied cooling rate. Herein, the ``ideal glass transition" that occurs at the finite Kauzmann temperature at which point the configurational entropy of an undercooled liquid matches that of its crystalline counterpart is identified as a first-order quantum critical point. We suggest that this quantum critical point belongs to quaternion ordered systems that exist in four- and three-dimensions. The Kauzmann quantum critical point is considered to be a higher-dimensional analogue to the superfluid-to-Mott insulator quantum phase transition, in two- and one-dimensional complex ordered systems. Such quantum critical points are driven by tuning a non-thermal frustration parameter, and result due to characteristic softening of the `Higgs' type mode that corresponds to amplitude fluctuations of the order parameter. The first-order nature of the finite temperature Kauzmann quantum critical point is seen as a consequence of the discrete change of the topology of the ground state manifold that applies to crystalline and non-crystalline solid states.
\end{abstract}
\maketitle

\section{Introduction}

The nature of the glass transition is widely thought to be one of the most challenging problems in condensed matter physics~\cite{anderson_through_1995, angell_insights_2008, mauro_statistical_2014, charbonneau_glass_2017}. Despite their ubiquity, a fundamental understanding of amorphous solids and the glass transition has yet to be obtained. This is in stark contrast to our knowledge of crystalline solids, which result by first-order phase transitions and in which phonons (collective Nambu-Goldstone modes) develop to transport thermal energy. Unlike crystallization, glass formation is non-equilibrium and the glass transition results as an undercooled liquid breaks ergodicity to become a rigid solid. 

The glass transition is highly dependent on the cooling rate; with a slower cooling rate, an undercooled liquid may remain ergodic to lower temperatures. As the temperature of an undercooled liquid is decreased, the difference in entropy between the liquid and crystalline solid phase decreases. Because the entropy of undercooled fluids declines faster with temperature than does crystal entropy, this results in an entropy paradox at a certain finite temperature that may be achieved in the limit of an infinitely slow cooling rate~\cite{kauzmann_nature_1948, stillinger_kauzmann_2001, speedy_kauzmanns_2003}. At this finite temperature, known as the Kauzmann temperature $T_K$, the liquid adopts a unique state of lowest energy.

An ``ideal glass," that forms at the Kauzmann temperature, has a configurational entropy that matches its crystalline counterpart. This is known in the literature  as the Kauzmann entropy paradox~\cite{kauzmann_nature_1948, stillinger_kauzmann_2001, speedy_kauzmanns_2003}. ``Ideal glass transitions" have never been obtained in the laboratory, as any real glass transition occurs at a temperature higher than $T_K$ and is caused by kinetic constraints that are responsible for a loss of ergodicity~\cite{debenedetti_supercooled_2001}. Yet, the physics underlying the Kauzmann entropy paradox presents an interesting problem in condensed matter physics.

The objective of this article is to describe the nature of formation of both crystalline and glassy solids from the liquid state, and the origin of the Kauzmann point, by the application of a quaternion orientational order parameter. This approach makes use of the principles of spontaneous symmetry breaking and of topological-ordering, that are known to play key roles in our understanding of condensed matter. Analogies between solidification and the formation of complex ordered states of matter, in two- and one-dimensions, are developed in order to frame the Kauzmann point as a quantum critical point (QCP) that exists to separate crystalline and glassy solid states. The topological properties of the free energy functions that apply to crystalline and glassy solids are compared, and the roles of the two types of fundamental exciations: massless phase modes (Nambu-Goldstone) and a massive amplitude mode (`Higgs') are discussed. These considerations enable the characterization of the ``ideal glass transition," at the finite Kauzmann temperature, as a first-order QCP similar to the second-order superfluid/Mott-insulator quantum phase transition (QPT) that occurs in two- and one-dimensions~\cite{endres_higgs_2012}.

\section{Crystalline-to-Glass First-Order Quantum Critical Point and the Kauzmann Entropy Paradox}

Quantum phase transitions~\cite{sachdev_quantum_2011} exist for $O(N)$ quantum rotor models of $N-$vector ordered systems, which are constructed by taking into account both potential and kinetic energy terms. Such QPTs~\cite{sachdev_quantum_2011} are driven by tuning a dimensionless frustration parameter ($g$), that is a measure of the ratio of kinetic energy to potential energy. Such quantum rotor models have been applied extensively to understand the nature of the superfluid/Mott-insulator QPT in two- and one-dimensions~\cite{endres_higgs_2012}. 

\begin{figure}[t!]
\centering
\includegraphics[scale=.6]{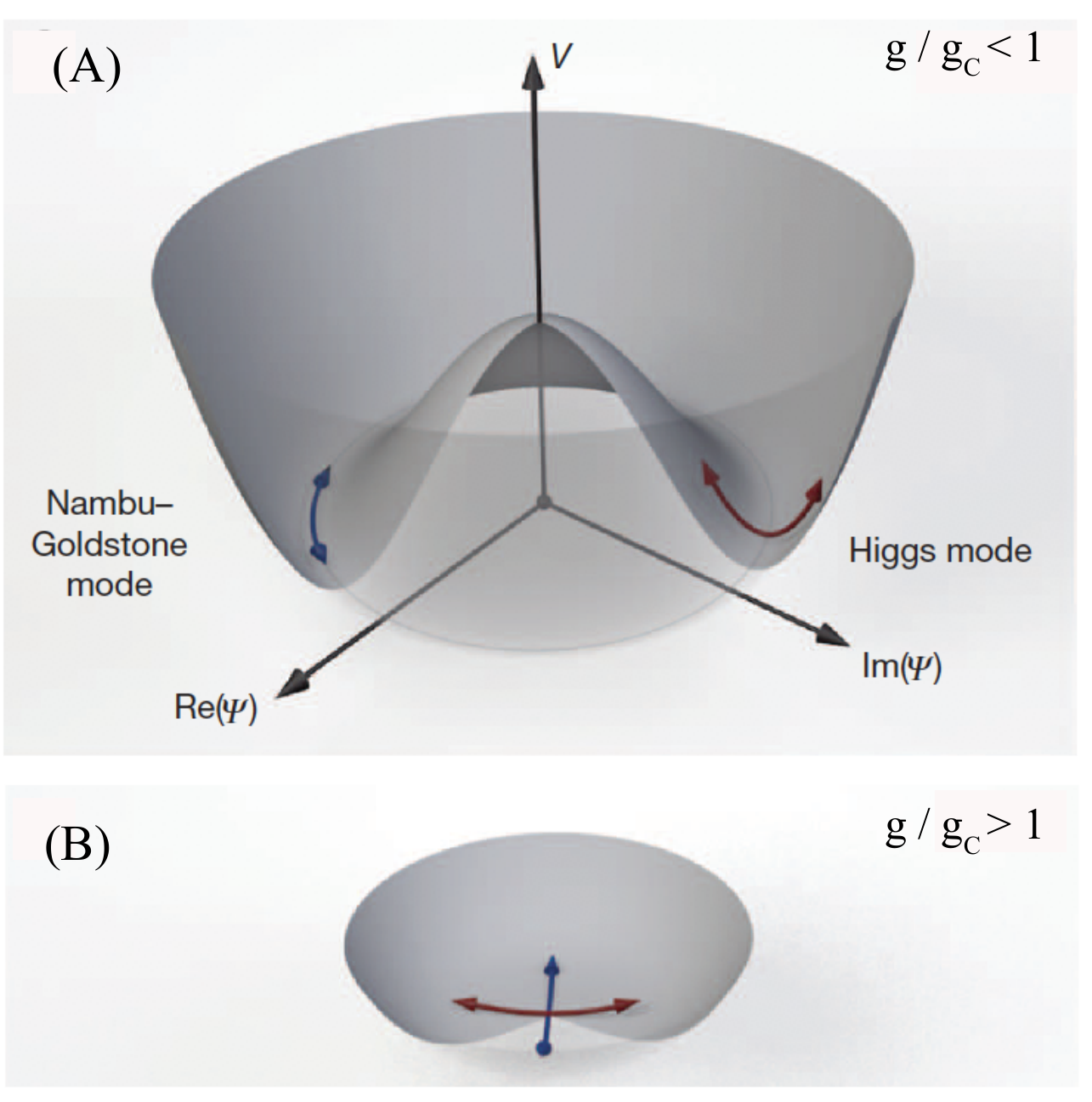}
\caption{(A) The `Mexican hat' free energy function of complex ordered systems whose complex order parameter has the form $\Psi=|\Psi|e^{\hat{i}\theta}$. In the phase-coherent superfluid phase, $|\Psi|>0$ and a massless Nambu-Goldstone and a massive Higgs modes arise. (B) On approaching the two-dimensional superfluid/Mott insulator QPT, at a critical value of frustration~\cite{endres_higgs_2012} ($g_C$), the free energy function transforms to one with a minimum at $|\Psi|=0$. [Reproduced from Ref.~\onlinecite{endres_higgs_2012}]}
\label{fig:mexicanhat}
\end{figure}

\begin{figure}[t!]
\centering
\includegraphics[scale=.4]{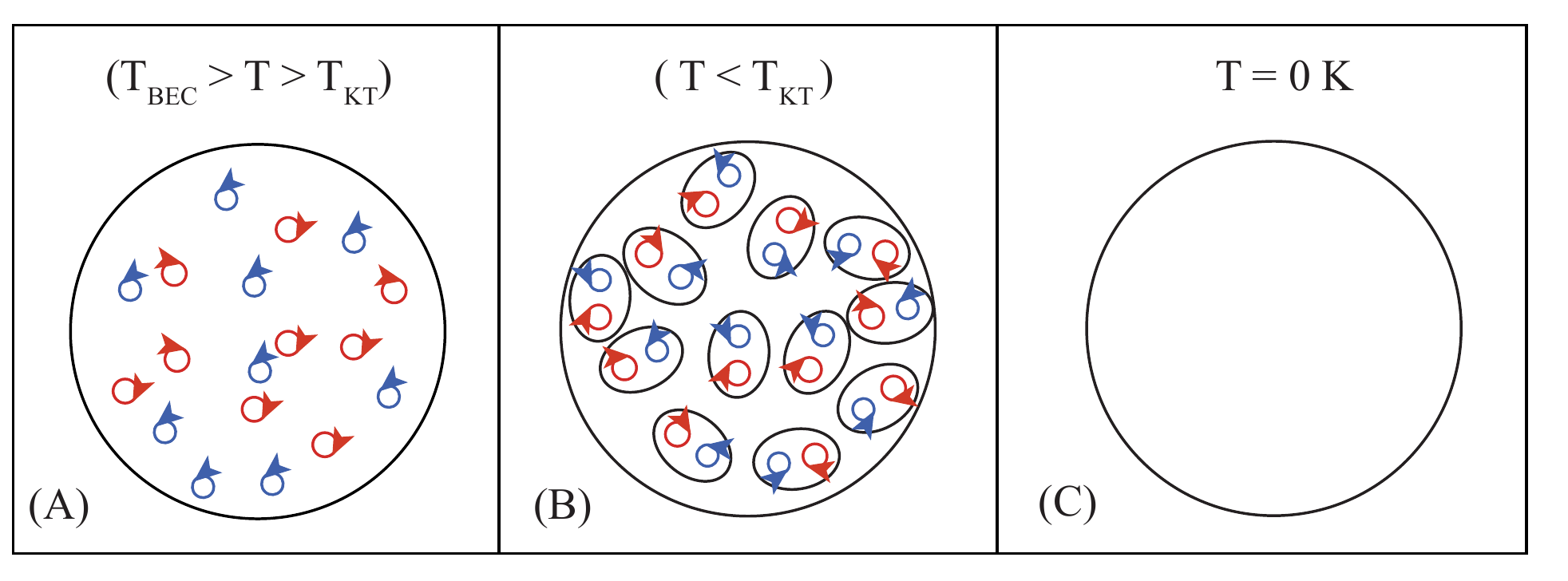}
\caption{Classical 2D $O(2)$ rotor model. (A) An abundance of misorientational fluctuations develops below the bulk Bose-Einstein condensation temperature ($T_{BEC}$), and may be discretized as a plasma of isolated point defects and anti-defects. (B) As the temperature is lowered below the Kosterlitz-Thouless transition temperature ($T_{KT}$), complementary defects/anti-defects begin to form bound pairs. (C) As the temperature approaches 0 K, defects and anti-defects that comprise bound states come together and annihilate. In the absence of frustration, no signed defects persist to the ground state that is perfectly phase-coherent.}
\label{fig:JJA_binding}
\end{figure}

In two- and one-dimensional complex ordered systems, an order parameter with $N=2$ components develops and $O(2)$ quantum rotor models describe the dynamics of the complex order parameter ($\Psi=|\Psi|e^{\hat{i}\theta}$) near a QPT between a phase-coherent superfluid ($|\Psi|>0$) and a phase-incoherent Mott-insulator ($|\Psi|=0$). In the superfluid ordered state of matter, the free energy function has the form of a conventional `Mexican hat' (Figure~\ref{fig:mexicanhat}A) where the order parameter has a non-zero value at the basin of the `Mexican hat.' In two- and one-dimensional complex ordered systems, as opposed to bulk three-dimensional complex ordered systems, phase-coherent superfluid ground states are achieved via a Kosterlitz-Thouless topological-ordering transition in which vortex defects and anti-defects form bound states (Figure~\ref{fig:JJA_binding}). In the absence of kinetic energy effects, i.e., in the absence of frustration, the scalar phase angle $\theta$ acquires a definite value below the Kosterlitz-Thouless transition through breaking of rotational symmetry -- and the superfluid ground state is perfectly phase-coherent. 

Around the symmetry-broken ground state, there are two mode types: a massless Nambu-Goldstone mode related to fluctuations in the scalar phase angle $\theta$ and a massive `Higgs' mode related to amplitude variations in $\Psi$. As the amount of frustration reaches a critical value, in the vicinity of the QPT to the phase-incoherent Mott-insulator, there is a characteristic softening of the excitation gap or mass of the Higgs amplitude mode~\cite{endres_higgs_2012}. This softening transforms the free energy into a function with a minimum at $|\Psi|=0$ in the phase-incoherent Mott-insulating state~\cite{endres_higgs_2012} (Figure~\ref{fig:mexicanhat}B).

\begin{figure*}[t!]
\centering
\includegraphics[scale=1]{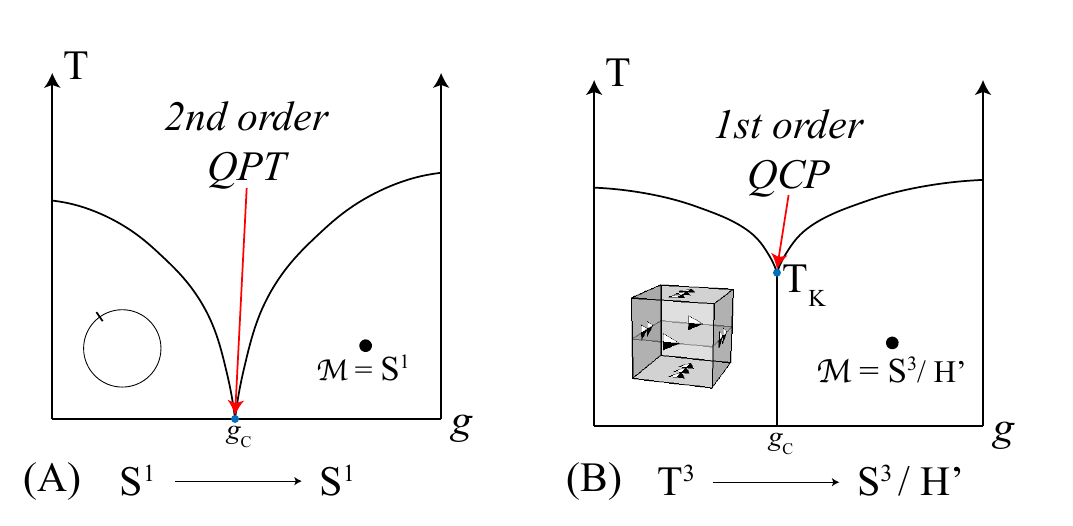}
\caption{(A) Complex ordered systems ($N=2$) that exist in 2D/1D are mathematically described using O($2$) quantum rotor models, that admit a second-order QPT at absolute zero temperature~\cite{baturina_superinsulatorsuperconductor_2013}. This is the superfluid/Mott-insulator QPT~\cite{endres_higgs_2012}. (B) Solidification processes in four- and three-dimensions, as characterized by a quaternion orientational order parameter ($N=4$), are described using $O(4)$ quantum rotor models. Such $O(4)$ quantum rotor models admit a QCP that is first-order. This may be identified with the  ``ideal glass transition,'' that occurs at a finite Kauzmann~\cite{kauzmann_nature_1948} temperature.}
\label{fig:tree_S1}
\end{figure*}

The order of the quantum phase transition, that belongs to an $O(N)$ quantum rotor model, can be discerned by noting changes in the topological properties of the free energy function in its vicinity. In the case of the superfluid/Mott-insulator QPT, the topology of the `Mexican hat' that applies to the superfluid is circular and the free energy of the Mott-insulator retains $U(1)\cong S^1$ symmetry at the origin. Thus, the superfluid/Mott-insulator QPT is continuous and is therefore of second-order. Such a second-order QPT occurs at zero Kelvin (Figure~\ref{fig:tree_S1}A), and describes a change in the ground state as a result of quantum fluctuations arising from the Heisenberg uncertainty principle~\cite{sachdev_quantum_2011}.

In the same way that the degree of order in superfluids (phase-coherent) is described by a complex order parameter, it has recently been proposed by the authors that orientational-order in crystalline solids may be described by a quaternion order parameter~\cite{gorham_topological_2019, gorham_crystallization_2019}. Quaternion order parameters $\Psi=|\Psi|e^{\hat{\tau}\phi}$, where $\hat{\tau}^2=-1$ and $\hat{\tau}=u_1\hat{i}+u_2\hat{j}+u_3\hat{k}$ is a unit imaginary quaternion vector, are four-dimensional and are described by a set of three scalar phase angle parameters. In four- and three-dimensions, $O(4)$ quantum rotor models apply to mathematically model quaternion ordered systems -- as these are ``restricted dimensions'' for quaternions in the Hohenberg-Mermin-Wagner sense~\cite{hohenberg_existence_1967, mermin_absence_1966, halperin_hohenbergmerminwagner_2018}. When applied to the solidification problem, a first-order quantum critical point is anticipated to exist between crystalline (i.e., orientationally-ordered) and non-crystalline (i.e., orientationally-disordered) solids.

Crystalline solids are anticipated to develop as a result of a defect-driven topological-ordering transition~\cite{gorham_su2_2018, gorham_crystallization_2019}, just as the Kosterlitz-Thouless mechanism~\cite{kosterlitz_ordering_1973, kosterlitz_critical_1974} allows for the realization of superfluids in 2D/1D. Just as first homotopy group defects are available to complex ordered systems (i.e., $\pi_1(S^1)$ vortices), third homotopy group defects are available to quaternion ordered systems ($\pi_3(S^3)$). Vortices are points in two-dimensions (complex plane) and, third homotopy group defects are points in four-dimensions (quaternion plane). In these ``restricted dimensions" (Hohenberg-Mermin-Wagner theorem~\cite{mermin_absence_1966, hohenberg_existence_1967, halperin_hohenbergmerminwagner_2018}), defect binding via a Kosterlitz-Thouless mechanism is necessary to prevent the mobility of misorientational fluctuations such that phase-coherency or long-range orientational-order may be obtained.

In addition to third homotopy group defects, closed-loop (fundamental group) defects~\cite{mermin_homotopy_1978,mermin_topological_1979, nelson_liquids_1983} exist below the melting temperature as a consequence of the discrete orientational-order in clustered undercooled fluids. Such defects are known as disclinations. Just as complementary third homotopy group point defects form bound pairs on crystallization, disclinations of equal and opposite sign come together to form dislocations~\cite{pretko_fracton_2018, halperin_theory_1978, chaikin_principles_2000, yazyev_polycrystalline_2014}. In perfect crystals, at absolute zero temperature, definite values for the set of three scalar phase angle parameters that define the quaternion order parameter are obtained as components of bound pair excitations come together and annihilate.

\begin{figure}[b!]
  \centering
\includegraphics[scale=.6]{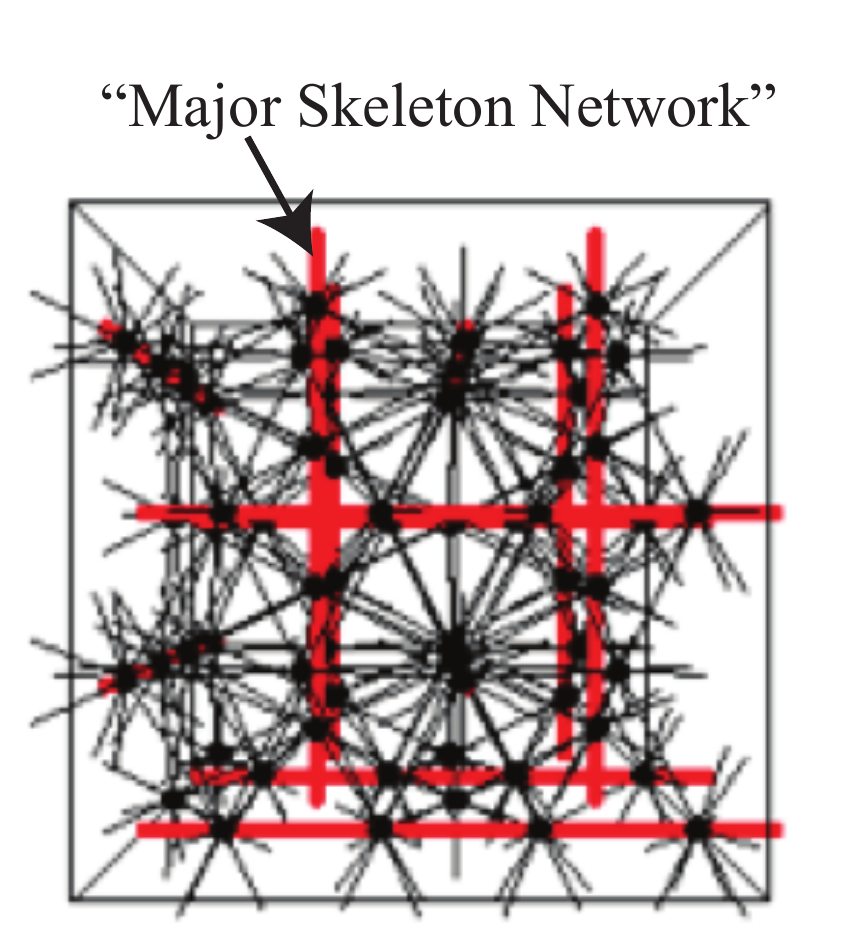}
\caption{In geometrically-frustrated crystalline structures, i.e., topologically close-packed, frustration-induced topological defects form a periodic arrangement. The ordered arrangement of negative wedge disclinations~\cite{frank_complex_1958, frank_complex_1959} is known as the ``major skeleton network." Signed third homotopy group defects also form a periodic arrangement, in geometrically-frustrated crystalline solids, but are not visible because of their nature as points in four-dimensions. [Reproduced from Ref.~\onlinecite{doye_effect_1996}]}
\label{fig:periodic_tangled}
\end{figure}

In the presence of finite frustration effects, on approaching the quantum critical point from the limit of a perfect crystal, crystalline solid states may form in which local orientational-order is incompatible with long-range crystallographic packing. Examples of such structures are topologically close-packed Frank-Kasper crystalline solids~\cite{frank_complex_1959, frank_complex_1958}, in which geometric frustration~\cite{nelson_symmetry_1984, sadoc_geometrical_2006} prevents the development of long-range icosahedral orientational-order. Such geometrically frustrated crystalline structures (e.g., Frank-Kasper) are stabilized in the ground state by a periodic arrangement of signed topological defects. In particular, the frustration-induced signed disclination lines that are present carry negative curvature and form what is known as a ``major skeleton network" (Figure~\ref{fig:periodic_tangled}). An entangled array of negative disclination lines is an attractive model for the structure of glasses~\cite{nelson_liquids_1983, nelson_symmetry_1984}, that form above a critical amount of geometrical frustration. Just as topologically close-packed crystalline solids may be viewed as analogous to 2D/1D superfluids with a finite amount of frustration, orientationally-disordered glasses are similar to the phase-incoherent Mott-insulator state~\cite{endres_higgs_2012}.

In crystalline solids, the order parameter manifold has the topology of a three-dimensional torus ($T^3\cong S^1\times S^1\times S^1$) that accommodates periodic boundary conditions. This can be viewed as a higher-dimensional, quaternionic, version of the `Mexican hat' of superfluids (Figure~\ref{fig:mexicanhat}A). Around the symmetry-broken crystalline ground state: three massless phonon modes (Nambu-Goldstone) and a massive mode (`Higgs'), related to the amplitude variations in $\Psi$, exist.

Like the 2D/1D superfluid/Mott-insulator QPT~\cite{endres_higgs_2012}, softening of the excitation gap of the amplitude mode is anticipated as frustration is increased to approach the QCP that belongs to $O(4)$ quantum rotor models. This softening transforms the order parameter manifold into a function with a minimum at $|\Psi|=0$, at the QCP, which retains $SO(3)/H\cong SU(2)/H'$ symmetry at the origin (where $H'$ is the binary polyhedral group of preferred local orientational order $H\in SO(3)$, e.g., $H'=\mathbb{Z}_2$ in the case of the trivial group that applies to continuous random network glasses~\cite{rivier_gauge_1990}). 

Owing to the discontinuous change in the genus topological invariant of the ground state manifold that applies to crystalline and glassy solids that the frustration-induced QCP is first-order, and thereby occurs at a finite temperature (Figure~\ref{fig:tree_S1}B). When considering solidification processes, the first-order QCP that belongs to $O(4)$ quantum rotor models may be identified as the Kauzmann point~\cite{kauzmann_nature_1948} that occurs at the finite Kauzmann temperature ( ``ideal glass transition''). At the Kauzmann temperature, a Kauzmann entropy paradox~\cite{} occurs. This entropy paradox, where the configurational entropy of an undercooled liquid and its crystalline counterpart are equal, is physically acceptable at finite temperatures but would not be so at the absolute zero of temperature. Although the finite temperature nature of the ``ideal glass transition'' is well-understood based upon thermodynamic principles, this topological interpretation of the first-order nature of this QCP is novel.

\section{Conclusions and Outlook}

The abundance of literature that exists on the nature of both the real and the ``ideal glass transition" clearly reflects the importance of the free energy landscape in providing a qualitative explanation of the phenomenon. In this article, we have suggested that crystallization and glass formation can be understood within a unified framework by the application of a four-dimensional quaternion orientational order parameter. As a generalization of the superfluid/Mott-insulator quantum phase transition (QPT) in two- or one-dimensions, a quantum critical point (QCP) is anticipated for quaternion ordered systems that  exist in four- or three-dimensions. This QCP has been identified with the ``ideal glass transition,'' that occurs at the finite Kauzmann temperature.
 
The first-order nature of the Kauzmann QCP has been determined by accounting for the discrete change in the topology of the ground state manifold that applies to crystalline (orientationally-ordered) and non-crystalline (orientationally-disordered) solid states. Just as in the case of the superfluid/Mott-insulator QPT, the ground state manifold is anticipated to become modified due to characteristic softening of the amplitude mode ('Higgs') on approaching the Kauzmann QCP from the limit of a perfect crystal. The first-principles approach presented in this article suggests that the Kauzmann temperature, at which the Kauzmann entropy paradox may be realized, is finite and thereby corresponds with experiments.

\section{Acknowledgment}

The authors acknowledge support from the ALCOA Chair in Physical Metallurgy. Additionally, CSG is grateful for insightful discussions with Prof. Mingda Li (MIT) and Prof. Di Xiao (CMU).

\bibliography{\jobname}

\end{document}